%% file: _main.tex
\PassOptionsToPackage{table}{xcolor}
\documentclass[sigconf]{acmart}

\usepackage{hyperref}
\usepackage{url}
\usepackage{graphicx}
\usepackage{subcaption}
\usepackage{multirow}
\usepackage{makecell}
\usepackage{tabularx}
\usepackage{balance}
\usepackage{enumitem}
\usepackage{pifont}
\usepackage{booktabs}
\usepackage{tikz}
\usetikzlibrary{positioning, arrows.meta, calc}
\definecolor{gold}{RGB}{255, 215, 0}
\newcommand{\mstd}[2]{\textbf{#1}{\scriptsize(#2)}}

\AtBeginDocument{%
  }

\begin{document}
\title[Codebook-Based Semantic IDs in Generative Recommendation]{Codebook-Based Semantic IDs in Generative Recommendation:\\Enabling Interface, Emerging Bottleneck}

\author{Danil Gusak}
\orcid{0009-0008-1238-6533}
\affiliation{%
  \institution{AXXX, AI Center}
  \city{Moscow}
  \country{Russian Federation}
}
\email{gusak.di18@physics.msu.ru}

\author{Evgeny Frolov}
\orcid{0000-0003-3679-5311}
\affiliation{%
  \institution{AXXX, HSE University}
  \city{Moscow}
  \country{Russian Federation}
}

\input{sections/00_rights}

\begin{teaserfigure}
\setlength{\abovecaptionskip}{6pt}
\centering
\includegraphics[width=1.0\textwidth]{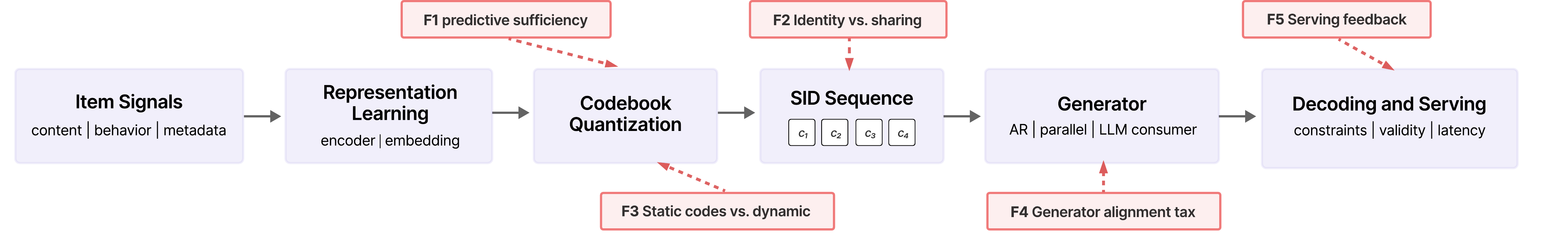}
\caption{Codebook-based semantic IDs act as a three-way contract between item representation, sequence generation, and runtime serving. The interface breaks when it fails to preserve task-relevant signal (F1), balance sharing against item specificity (F2), remain valid under drift (F3), align with the consumer model (F4), or execute within serving constraints (F5).}
\label{fig:pipeline}
\end{teaserfigure}

\begin{abstract}
\input{sections/00_abstract}
\end{abstract}

\maketitle

\section{Introduction}\label{sec:intro}
\input{sections/01_introduction}

\section{Past: How Canonical Two-Stage Pipeline
Won}\label{sec:past}
\input{sections/02_sec2}

\section{Present: Five Fractures in the Pipeline}\label{sec:present}
\input{sections/03_sec3}

\section{Benchmarking and Evidence Quality}\label{sec:bench}
\input{sections/04_benchmarking}

\section{Future: From Semantic Compression to
Identity Contracts}\label{sec:future}
\input{sections/05_future}

\section{Conclusion}\label{sec:conclusion}
\input{sections/06_conclution}

\bibliographystyle{ACM-Reference-Format}
\balance
\bibliography{bibliography}

\end{document}

%% file: sections/00_rights.tex
%% Rights management information.  This information is sent to you
%% when you complete the rights form.  These commands have SAMPLE
%% values in them; it is your responsibility as an author to replace
%% the commands and values with those provided to you when you
%% complete the rights form.
\copyrightyear{2026}
\acmYear{2026}
\setcopyright{cc}
\setcctype{by}
\acmConference[RecSys '26]{20th ACM Conference on Recommender Systems}{September 27-October 02, 2026}{Minneapolis, MN, USA}
\acmBooktitle{20th ACM Conference on Recommender Systems (RecSys '26), September 27-October 02, 2026, Minneapolis, MN, USA}
\acmDOI{10.1145/3773078.3831806}
\acmISBN{979-8-4007-2284-4/2026/09}

\begin{CCSXML}
<ccs2012>
   <concept>
       <concept_id>10002951.10003317.10003347.10003350</concept_id>
       <concept_desc>Information systems~Recommender systems</concept_desc>
       <concept_significance>500</concept_significance>
       </concept>
 </ccs2012>
\end{CCSXML}

% \ccsdesc[500]{Information systems~Recommender systems}
\ccsdesc[500]{Information systems~Recommender systems}

\keywords{recommender systems; semantic IDs; generative recommendation; item tokenization; serving constraints}

%% file: sections/00_abstract.tex
Codebook-based semantic IDs (SIDs), short discrete code sequences produced by quantization over item representations, made generative recommendation practical by turning catalog-scale retrieval into low-cardinality generation. Yet the same interface now concentrates the field's hardest questions. We revisit the canonical two-stage pipeline for SID-based recommendation --- item tokenization using RQ-VAE, K-means variants, or related methods, followed by training a generator to emit the resulting codes --- as a past--present--future story. Our thesis is that these identifiers succeeded as an enabling interface, but not yet as a stable theory of item identity for recommendation: a single SID must simultaneously preserve behaviorally useful similarity, protect identity under head-tail skew, survive catalog drift, align with language-model generators, and remain decodable under production constraints. We organize the present literature around five recurring fractures --- objective mismatch, identity-versus-sharing tension, static codes in dynamic settings, LLM alignment tax, and serving feedback into identifier design --- and argue that SID utility is regime-dependent: model capacity, catalog scale, and interaction density jointly determine whether semantic structure helps or is redundant. We close with a research agenda centered on recommendation-native tokenization, adaptive identity, explicit alignment protocols, and co-design of tokenization with serving.

%% file: sections/01_introduction.tex
Generative retrieval and large language models revived an old ambition in recommender systems: replace explicit scoring over massive catalogs with sequence generation over a learned item vocabulary.
Codebook-based semantic IDs (SIDs) made that ambition practical~\cite{tiger}.
Instead of predicting a raw item ID from an effectively unstructured output space, early generative recommenders first mapped each item into a short sequence of discrete codes and then trained a model to generate that sequence autoregressively~\cite{tiger}.
This move reduced output cardinality, exposed shareable structure across items, and made recommendation legible to the tooling of modern generative modeling.

This paper focuses on a specific lineage within that broader space.
By \emph{codebook-based semantic IDs} we mean learned discrete code sequences produced by quantization over item representations: RQ-VAE-style tokenizers, residual or balanced K-means variants, and related differentiable descendants~\cite{vqvae,tiger}.
Throughout, the object of analysis is the \emph{canonical two-stage pipeline}: tokenize items offline, then train a generator to emit those codes.
We treat textual identifiers, term-based identifiers, and dense retrieval as boundary cases that pressure-test this pipeline rather than as its core.
The key question is not ``what methods exist?'', but ``what has this influential interface actually taught us, and where is it now breaking?''

Three years in, the answer is surprisingly unsettled.
Simple tokenizers match complex ones on academic benchmarks~\cite{grid}, and strong non-SID baselines can outperform reported SID methods under common offline evaluation protocols. Yet SID-based production systems report genuine online gains~\cite{onerec,static,gpr,onemall,sigma}.
A useful lens for resolving these contradictions is to view an SID not as a semantic description of an item, but as a \emph{contract between three layers}: the representation that produces the code, the generator that consumes it, and the serving stack that validates it.
Much of the recent literature can be read as discovering where that contract is underspecified.

Our claim is simple: codebook-based SIDs were the interface innovation that made generative recommendation plausible, but they are now also where the field's unresolved assumptions concentrate.
One identifier is asked to do too many jobs: preserve recommendation-relevant similarity, protect identity under head--tail skew, survive catalog drift, align with generators that may not be language-native, and satisfy latency and validity constraints.
The recent literature is best read not as a pile of unrelated tricks, but as repeated repairs to the same few stresses.
Our contributions are threefold: (i) a diagnostic framework of five recurrent fracture points in the two-stage SID pipeline, (ii) a descriptive audit of evaluation practice and released baselines, and (iii) a regime-aware agenda for future identifier design.
% Section~2 revisits why conventional SIDs won, Section~3 diagnoses five fractures, Section~4 describes the problems in benchmarking SID-based recommenders, and Section~5 outlines a future agenda.

%% file: sections/02_sec2.tex
Historically, codebook-based SIDs emerged at the intersection of three ideas.
The first was discrete latent modeling: VQ-VAE showed that learned codebooks could serve as compact discrete interfaces for downstream generation~\cite{vqvae}.
The second was residual quantization: stacking codebooks in sequence made short discrete codes practical without collapsing representational quality~\cite{rqvae,tiger}.
The third was model-based indexing from information retrieval: DSI and NCI showed that retrieval itself could be reframed as identifier generation rather than explicit nearest-neighbor search~\cite{dsi,nci}.
TIGER transferred that logic into recommendation and crystallized the canonical recipe: build an SID offline, then train a sequence model to generate the next SID from interaction history~\cite{tiger}.

This was more than an implementation detail.
Codebook-based SIDs offered a compromise between semantically empty raw IDs and full-text generation.
They reduced catalog complexity, opened a path for cold-start transfer through content-aware code sharing~\cite{damico2026glide,musicsid,coins}, and made recommendation compatible with the rapidly growing generative modeling stack~\cite{wang2024bettergen,sidsurvey}.
Multiple industrial systems adopted the paradigm independently --- Kuaishou~\cite{onerec,gems,zhou2025onerecv2,openonerec,onemall}, Taobao~\cite{forge}, Tencent~\cite{gpr}, YouTube~\cite{static, plum}, Spotify~\cite{damico2026glide}.

But that success also locked in assumptions that later work has been systematically testing: that one mostly stable code per item is desirable, that reconstruction or generic semantic faithfulness is a good surrogate for recommendation utility~\cite{cost,resid,colarec}, that tokenization can be cleanly separated from generation~\cite{diger,etegrec,cofirec}, and that serving concerns mostly begin after model design~\cite{static,zhou2025onerecv2}.
This historical framing matters because it explains both the field's initial speed and its current fragility: recommendation inherited powerful tools from neighboring areas --- VQ-VAE from vision~\cite{vqvae}, hierarchical indexing from IR~\cite{dsi,nci} --- but not objectives originally designed around user-behavior prediction.

%% file: sections/03_sec3.tex
The current literature is best understood not as a growing inventory of methods but as repairs to five fractures in the conventional pipeline (Figure~\ref{fig:pipeline}) --- persistent tensions that explain why the field is simultaneously innovative and fragmented. These fractures are coupled rather than independent: what the tokenizer preserves affects collisions, static codes amplify alignment problems when catalogs drift, and serving constraints feed back into what counts as a viable identifier at all. Which fracture binds most strongly also appears to depend on regime: small academic systems care more about representational structure, while large production systems increasingly surface alignment and serving costs. Another recurring pattern is displacement rather than resolution: methods that ease one fracture often intensify another. More adaptive tokenization complicates alignment and serving, stronger collision repair can protect identity while reducing useful sharing, and stricter constrained decoding can rescue validity while shrinking the practical benefit of richer code spaces.

This regime dependence also helps explain why the literature can look contradictory. On smaller academic benchmarks, almost any reasonably distinctive code often helps because the main problem is giving a modest model enough structure to avoid memorizing arbitrary item mappings. At industrial scale (large models, rich YouTube-scale~\cite{plum, static} histories), the bottleneck shifts: simple codes may already be expressive enough, and the harder questions become alignment cost, dynamic maintenance, and decoding under latency. That is why some papers report surprisingly strong simple or even hash-like baselines, while deployed systems invest heavily in grounding, constrained decoding, and index maintenance~\cite{wang2024bettergen,damico2026glide,zhou2025onerecv2,oxygenrec,sigma,gr4ad}. The disagreement is therefore often not about whether SIDs matter, but about which operating regime a result is speaking to.

\subsubsection{F1: What should a SID preserve?}
The first fracture is objective mismatch.
Conventional pipelines quantize embeddings optimized for reconstruction or semantic similarity, but recommendation needs \emph{predictive sufficiency} --- the information sufficient for next-item prediction, ranking, and transfer, which may be a strict subset of what reconstruction preserves.

CoST~\cite{cost} and SimCIT~\cite{simcit} replace reconstruction with contrastive losses: SimCIT removes the decoder entirely and proves convergence to a minimal sufficient identifier.
LETTER~\cite{letter} and COSETTE~\cite{cosette} inject collaborative signals, ReSID~\cite{resid} and DIGER~\cite{diger} make tokenization end-to-end differentiable.
An information-theoretic view frames the correct objective as a task-aware bottleneck, but no paper operationalizes this with the recommendation target as the relevance variable~\cite{diger}.
DIGER~\cite{diger} proves this matters formally: the gap between two-stage optimization (train tokenizer, then freeze and train generator) and joint optimization can be made \emph{arbitrarily large}, yet naive end-to-end training through discrete codes collapses codebooks catastrophically~\cite{e2esid}.
The community is therefore stuck between a provably suboptimal two-stage pipeline and a theoretically superior but practically fragile end-to-end one.
Meanwhile, it still infers ``good tokenization'' only \emph{after} training a full recommender, a costly indirection that blocks principled tokenizer design and makes the field unable to tell whether tokenization or generation is the binding constraint.

\subsubsection{F2: When does sharing become identity loss?}
The second fracture is the tension between useful sharing and item identity. Shared codewords are precisely why SIDs help cold-start and long-tail items: they induce parameter sharing across related neighborhoods. But the same mechanism creates collisions, codebook imbalance, and head--tail trade-offs. Recent work on qualification-aware collision repair and variable-length SIDs makes explicit that collisions are not uniformly bad~\cite{hu2026quasid,khrylchenko2026varlen,intrr,mmgrec}. A harmful collision erases a distinction the recommender needs, a benign collision can transfer signal where data are sparse. This suggests a useful interpretive lens: conventional SIDs are not only compressed semantics. In practice they often behave like \emph{structured collaborative hashes}, whose value lies in making the right items collide for the right reasons. That lens is consistent with ranking evidence that semantic IDs improve generalization over random hashes in large, dynamic catalogs~\cite{wang2024bettergen}. The design question is therefore not how to eliminate all collisions, but which sharing patterns should be encouraged, for whom, and at what catalog frequency.

\subsubsection{F3: Can one fixed SID survive dynamic recommendation?}
The canonical pipeline assigns each item one code built offline.
But catalogs grow, collaborative structure drifts~\cite{baikalov2026mitigating}, and item meaning depends on context.
MERGE~\cite{merge} constructs codes from streaming data,
Pctx~\cite{zhong2025pctx} and PIT~\cite{wang2026pit} make tokenization user-dependent or co-evolving with
the recommender,
ActionPiece~\cite{hou2025actionpiece} tokenizes items differently based on behavioral context.
Yet dynamic codes conflict with caching, constrained decoding, and serving efficiency.
The tension is structural: static codes are easy to index, cache, and validate via prefix tries, but recommendation operates over drifting distributions where yesterday's codebook is today's approximation.
Item identity in recommendation is partly relational and temporal, conventional pipelines have only begun to absorb this, and it remains unclear how quickly static codes become stale in production.

\subsubsection{F4: The LLM alignment tax} The fourth fracture is alignment. SID tokens are foreign to the native vocabulary of language models. Their embeddings must be initialized, grounded, and adapted through staged training, while practitioners decide what to freeze, when to unfreeze, and which auxiliary tasks best teach token semantics~\cite{minionerec,lcrec,openonerec}. Recent work on semantic-aware initialization and bidirectional SID--text alignment makes this explicit, and production systems such as Spotify's podcast recommender devote a distinct grounding stage to teaching the model how to ``speak catalog''~\cite{feng2026tsrec,damico2026glide,sigma}. An alternative response is to make identifiers more language-like, as in structured term identifiers, so the generator can reuse its native vocabulary~\cite{zhang2026grlm,idgenrec,lmindexer}. The point here is not to pick a winner yet, it is that alignment remains a tax whose cost is rarely isolated or reported cleanly.

The tax is also deeper than vocabulary expansion. NLP tokenizers are judged partly by sequential predictability: good tokens make left-to-right modeling easier because later units are informative but not arbitrarily unpredictable given the prefix. Most SID tokenizers, by contrast, are optimized for reconstruction fidelity or semantic distortion, objectives that are agnostic to whether deeper code levels are easy for an autoregressive model to predict. A code can therefore be excellent as a quantizer yet awkward as a sequence. Seen this way, some alignment failures are really consumer--producer mismatches: the tokenizer is optimized for one downstream consumer, while the generator behaves like another.

\subsubsection{F5: When serving constraints reshape the interface}
The fifth fracture is operational.
Coarse-to-fine autoregressive SID decoding makes early token decisions effectively irreversible: once the prefix is wrong, later refinement cannot recover the lost candidate set.
This irreversibility has spawned a chain of RL patches (DPO, GRPO, process-reward-guided search~\cite{onerec,promise,gpr,softmaxdpo,rankgr}) each fixing one failure mode (popularity bias, semantic drift) while introducing another (instability, reward hacking, exponential search cost).
Parallel~\cite{rpg} and diffusion-based~\cite{lladarec,mdgr} methods address the structural problem rather than patching symptoms.
Meanwhile, beam width and generation strategy~\cite{grace, promise,volodkevich2025autoregressive}, constrained decoding~\cite{grid}, and validity checks interact directly with identifier design. STATIC shows
how much engineering is required just to make trie-based constraints accelerator-friendly~\cite{static}, while OneRec-V2~\cite{zhou2025onerecv2} shows that generation architecture can dominate the compute budget in large-scale
systems. Joint search-recommendation systems and hybrid sparse-dense pipelines suggest SIDs may work best as an exploratory interface that narrows candidate space before denser components take over~\cite{cobra,gensar}. The lesson is that serving is not an afterthought. Latency budgets, constraint management, and decoding
strategy feed back into what counts as a viable SID in the first place.

This changes how identifier quality should be framed.
A richer code is not automatically better if it requires very wide beams, brittle constraints, or repeated inference-time repair.
A slightly less expressive identifier may be the more pragmatic choice if it survives catalog churn, respects serving budgets, and composes with downstream rankers.
In other words, identifier quality is partly operational quality: measured not only by what information the code stores, but by what the system can reliably extract at run time.

%% file: sections/04_benchmarking.tex
Benchmarking is essential, yet our audit of 124 non-survey papers\footref{github} from 2023--2026 suggests that evaluation discipline has lagged behind modeling novelty. The first problem is protocol mismatch:  although 20\% of papers report online A/B tests, offline evaluation still relies on leave-one-out-style (LOO) splits (88\%), while only 12\% use global temporal splitting~\cite{timetosplit2025}. LOO is convenient, but it causes data leakage affecting accuracy of the evaluation results~\cite{timetosplit2025, sun2023take, ji2023critical}. The effect is amplified by a benchmark monoculture: academic results are concentrated on small, dense Amazon-2014 5-core subsets (Beauty, Sports, Toys)~\cite{mcauley2015amazon}, with catalogs typically in the $10^4$ range, whereas deployed systems operate at $10^7$--$10^9$ scale. Only two industrial systems publicly release benchmark data (FORGE~\cite{forge}, RecIF-Bench~\cite{openonerec}), so transfer from academic wins to production remains largely assumed rather than demonstrated.

The second problem is strong-baseline neglect. In recent SID papers, SASRec~\cite{kang2018sasrec} is either omitted or trained with BCE loss. Table~\ref{tab:sasrec_ce_audit} shows what happens when that single choice is corrected in the released pipelines we could reproduce, while keeping the rest of the setup unchanged: a plain SASRec with \emph{full cross-entropy loss} (CE) over the entire item catalog~\cite{klenitskiy2023turning, petrov2023gsasrec, mezentsev2024scalable, gusak2024rece} often outperforms popular SID methods, despite far lower modeling and configuration cost, making some reported \emph{SID gains} on small academic benchmarks not robust to a strong non-SID sequential baseline. The code for reproducing our experiments is available in the repository.\footnote{Code, paper list, inclusion criteria: \url{https://github.com/dalibra/sid-generative-recsys}\label{github}}

The third problem is evidentiary quality. Sixty percent of papers report neither significance tests, confidence intervals, nor multi-run variance, only 6.5\% report both significance and variance information. Only 40\% of papers release usable code or artifacts (code release rate for industrial papers  is 6$\times$ lower than for academic papers), and copying baseline numbers across papers despite changed preprocessing or metric code remains common. The risk is not only irreproducibility but misattribution: improvements caused by split choice~\cite{timetosplit2025, volodkevich2026splitlight}, weak baselines, or evaluation-code drift can be easily interpreted as progress in identifier design.

    \begin{table}[t]
        \setlength{\abovecaptionskip}{8pt}
        \caption{Amazon Beauty audit in public SID pipelines. For each pipeline, we report the original paper’s proposed method, the SASRec baseline from the paper, and our SASRec trained with \emph{full cross-entropy} (CE), keeping the evaluation setup unchanged. SASRec+CE reports 5-seed means, with standard deviations in parentheses on last shown digits.\footref{github}}
        \label{tab:sasrec_ce_audit}
        \centering
        \resizebox{\columnwidth}{!}{
            \begin{tabular}{llcccccc}
                \toprule
                Method & Metric & TIGER~\cite{tiger} & LETTER~\cite{letter} & RPG~\cite{rpg} & EAGER~\cite{eager} & LIGER~\cite{liger} & OpenOneRec~\cite{openonerec} \\
                \midrule
                \multirow{2}{*}{SASRec (paper)}
                & N@10 & 0.0318 & 0.0313 & 0.0318 & 0.0318 & 0.0218 & 0.0289 \\
                & R@10 & 0.0605 & 0.0588 & 0.0605 & 0.0605 & 0.0511 & 0.0639 \\
                \midrule
                \multirow{2}{*}{SASRec+CE}
                & N@10 & \mstd{0.0516}{2} & \mstd{0.0574}{8} & \mstd{0.0574}{8} & \mstd{0.0574}{8} & \mstd{0.0565}{5} & \mstd{0.0576}{6} \\
                & R@10 & \mstd{0.0875}{1} & \mstd{0.0930}{12} & \mstd{0.0930}{12} & \mstd{0.0930}{12} & \mstd{0.0928}{9} & \mstd{0.0937}{6} \\\midrule
                \multirow{2}{*}{Paper method}
                & N@10 & 0.0384 & 0.0418 & 0.0464 & 0.0525 & 0.0402 & 0.0545 \\
                & R@10 & 0.0648 & 0.0703 & 0.0809 & 0.0836 & 0.0745 & 0.0924 \\
                \midrule
                \rowcolor{gold!18}
                & N@10 & $-$25.51\% & $-$27.13\% & $-$19.11\% & $-$8.48\% & $-$28.87\% & $-$5.38\% \\
                \rowcolor{gold!18}
                \multirow{-2}{*}{vs. SASRec+CE} & R@10 & $-$25.96\% & $-$24.44\% & $-$13.04\% & $-$10.14\% & $-$19.75\% & $-$1.34\% \\
                \bottomrule
            \end{tabular}
        }
    \end{table}

%% file: sections/05_future.tex
The fractures above should be read not as indictments but as growing pains of an interface whose practical value is increasingly clear.
Industrial deployments consistently report genuine online gains, but the pattern is instructive:
the clearest gains are often reported in discovery- and cold-start-oriented settings~\cite{damico2026glide,static,musicsid,higen}, whereas improvements on warm-item ranking tasks are typically smaller~\cite{onerec,gpr,forge,onemall}.
This is consistent with the interpretation that SIDs' strongest realized value is \emph{exploration}, surfacing content that traditional ID-based systems cannot reach, not exploitation of well-observed items.
A mature deployment may benefit from positioning SID-based generation as a complementary discovery and cold-start channel that cooperates with dense or business-aware ranking components.

\textbf{Recommendation-native sufficiency.}
The next phase should move beyond reconstruction as the default objective for item tokenization and ask instead what information is \emph{sufficient} for behavior prediction, ranking, and transfer. A natural formalization is the information loss induced by discretization:
\begin{equation}\Delta I = I(X; Y \mid C) - I(Z; Y \mid C),\end{equation}
where $X$ is the continuous item representation, $Z$ the discrete SID, $C$ the user/context/history, and $Y$ the recommendation target.
To our knowledge, prior work has not directly estimated $\Delta I$, so the field cannot yet tell which side of the bottleneck it is optimizing.
Even coarse approximations (e.g., variational bounds on mutual information, or comparing the recommendation-quality ceiling of continuous embeddings against discrete codes on the same generator) would help determine whether the binding constraint lies in tokenization or in generation.
More broadly, tokenizer design should aim at recommendation utility (e.g., downstream ranking fidelity, cold-start transfer rate) rather than proxy quantities like reconstruction error or codebook utilization that may not correlate with recommendation performance.

\textbf{Adaptive and multi-view identity.}
One item need not map to one universally valid code. Recent work on contextual, personalized, variable-length, and co-evolving tokenization suggests that recommendation identity is layered~\cite{hou2025actionpiece,zhong2025pctx,wang2026pit,khrylchenko2026varlen,mbgen,cofirec}. A practical research path is therefore \emph{multi-view identity}: stable backbone codes where serving requires stability, augmented by contextual or scenario-specific refinements where the single-code assumption is clearly too weak.

\textbf{Alignment as protocol, not recipe.} If conventional SIDs remain part of LLM-based recommendation, the field needs explicit alignment protocols rather than a growing collection of local recipes. Papers should report how foreign their item tokens are, how embeddings are initialized, what is frozen or unfrozen, which auxiliary tasks are used to ground token meaning, and how much of the final gain is attributable to alignment rather than to the base recommender. Language-native alternatives make that burden of proof harder to avoid~\cite{zhang2026grlm,idgenrec}.

\textbf{Sequential predictability as a design objective.} If SIDs are consumed autoregressively, they should also be judged as sequences. Future tokenizer work should report whether deeper code levels remain predictable under a prefix, whether minor prefix errors are recoverable, and whether the identifier layout cooperates with the chosen decoder. This would move the field closer to the principle that made NLP tokenization effective: good units are not only compact but also easy for the model to continue.

\textbf{Tokenizer-decoder co-design.} Finally, tokenization and serving should be designed together. Variable-length codes, constrained decoding, accelerator-friendly constraints, and hybrid sparse--dense retrieval should be treated as coupled design choices rather than separate stages. In that view, an SID is not merely a compressed description of an item. It is a contract between a catalog, a generator, and a serving stack. The most plausible near-term role of codebook-based SIDs may therefore be narrower but clearer than early rhetoric suggested: they are especially valuable for candidate-space compression, exploration, and cold or sparse transfer, while dense or business-aware components may remain better suited for the final warm-item ranking pass.

\textbf{Regime-aware and benchmarking claims.} The binding fracture is unlikely to be universal. Small academic models with sparse interactions benefit from structured sharing differently than large production systems with rich histories and hard serving constraints. Future papers should therefore report the regime they speak to (at least model capacity, catalog scale, and interaction density~\cite{gusak2025warm}) before generalizing claims about identifier design.
They should also report strong baselines and benchmarking that can distinguish genuine advances in SID design from artifacts of protocol and infrastructure.

%% file: sections/06_conclution.tex
Codebook-based SIDs made generative recommendation practical by turning catalog-scale retrieval into manageable generation. But they have also become the point where the field's main unresolved questions now concentrate. A single identifier is asked to preserve predictive structure, retain identity, adapt to drift, align with generators, and serve efficiently.
The next step is to move from semantic compression toward recommendation-native identity design: identifiers judged by predictive sufficiency, adaptability, alignment cost, and operational viability. The real question is no longer whether semantic IDs are good or bad, but when they are the right interface, and how they should be designed for the regime they operate in.